\documentclass[conference]{IEEEtran}
\usepackage{amsfonts}
\usepackage{amssymb}
\usepackage{stfloats}
\usepackage{cite}
\usepackage{graphicx}
\usepackage{psfrag}
\usepackage{subfigure}
\usepackage{amsmath}
\usepackage{array}
\usepackage{algorithm}
\usepackage{algpseudocode}
\usepackage{setspace}
\usepackage[english]{babel}
\usepackage{blindtext}
\usepackage{url}
\usepackage{epstopdf}
\usepackage{verbatim}
\usepackage{color}
\usepackage{amsmath}
\usepackage{bm}
\usepackage{multirow}
\usepackage{booktabs}
\usepackage{makecell}
\newtheorem{Prob}{Problem}

\newcommand{\RNum}[1]{\uppercase\expandafter{\romannumeral #1\relax}}
\newcommand{\tabincell}[2]{\begin{tabular}{@{}#1@{}}#2\end{tabular}}

\title{Robust Sub-meter Level Indoor Localization - A Logistic Regression Approach}
\author{Chenlu Xiang$^{\dagger}$, Zhichao Zhang$^{\dagger}$, Shunqing Zhang$^{\dagger}$, Shugong Xu$^{\dagger}$, Shan Cao$^{\dagger}$ and Vincent LAU$^{\ddag}$\\
$^{\dagger}$ Shanghai Institute for Advanced Communication and Data Science, \\
Key laboratory of Specialty Fiber Optics and Optical Access Networks, \\
Shanghai University, Shanghai, 200444, China\\
$^{\ddag}$ Department of ECE, Hong Kong University of Science and Technology, \\
 Clear Water Bay, Kowloon, Hong Kong S. A. R., China \\
Email:\{xcl, zhichaozhang, shunqing, shugong, cshan\}@shu.edu.cn, eeknlau@ust.hk}

\begin{document}
\maketitle
\begin{abstract}
Indoor localization becomes a raising demand in our daily lives. Due to the massive deployment in the indoor environment nowadays, WiFi systems have been applied to high accurate localization recently. Although the traditional model based localization scheme can achieve sub-meter level accuracy by fusing multiple channel state information (CSI) observations, the corresponding computational overhead is significant. To address this issue, the model-free localization approach using deep learning framework has been proposed and the classification based technique is applied. In this paper, instead of using classification based mechanism, we propose to use a logistic regression based scheme under the deep learning framework, which is able to achieve sub-meter level accuracy (97.2cm medium distance error) in the standard laboratory environment and maintain reasonable online prediction overhead under the single WiFi AP settings. We hope the proposed logistic regression based scheme can shed some light on the model-free localization technique and pave the way for the practical deployment of deep learning based WiFi localization systems.
\end{abstract}
\begin{IEEEkeywords}
localization, deep learning, channel state information, logistic regression
\end{IEEEkeywords}

\section{Introduction} \label{sect:intro}
Indoor localization becomes a raising demand in our daily lives, which triggers a brand-new experience in the shopping mall and exhibition hall navigation \cite{chintalapudi2010indoor}. Different from the traditional outdoor localization where Global Navigation Satellite Systems (GNSS) \cite{marais2005land} is suffering from the satellite signal blocking effect, indoor localization often relies on a newly deployed infrastructure, such as infrared equipment \cite{hauschildt2010advances}, radio frequency identification (RFID) receivers and tags \cite{bouet2008rfid}, Bluetooth systems \cite{altini2010bluetooth}, sound or ultrasonic collection systems \cite{ijaz2013indoor}, or even hybrid of them. Despite complex indoor environment, high localization accuracy (within meter range) is still expected to offer satisfactory indoor location-based service.

WiFi systems, with the massive deployment in the indoor environment nowadays, have been applied to high accurate localization recently \cite{youssef2005horus,kotaru2015spotfi,vasisht2016decimeter,sen2012you,chapre2015csi}, since the additional deployment cost is usually marginal, if compared with aforementioned schemes. Among the existing WiFi localization techniques, fingerprints based scheme \cite{youssef2005horus} is shown to be a suitable solution, where we extract the features of indoor environments using WiFi signals in the training stage and predict the locations based on real time measured signals in the operating stage. A typical fingerprints based localization system is called ``HORUS'' \cite{youssef2005horus}, which utilizes the received signal strength indication (RSSI) as signal features. To further improve the localization accuracy, CSI has been proposed as useful WiFi signals in \cite{kotaru2015spotfi,vasisht2016decimeter,sen2012you,chapre2015csi}. In the propagation model based scheme, angle of arrival (AOA) and time of flight (TOF) information are often extracted from CSI to compute the relative location from the reference access points (APs), such as SpotFi \cite{kotaru2015spotfi} and Chronos \cite{vasisht2016decimeter}, while in the probability model based scheme, classifiers, such as deterministic k-nearest neighbor (KNN) clustering and probabilistic Bayes rule algorithms, are adopted to process the channel responses \cite{sen2012you,chapre2015csi}. The above model based solution can achieve sub-meter level accuracy if CSIs from multiple APs \cite{kotaru2015spotfi}, multiple frequency bands \cite{vasisht2016decimeter} or multiple antennas \cite{chapre2015csi} can be fused together. However, the corresponding computational overhead in the offline modeling and online feature extraction is quite significant \cite{wang2017biloc}.

In the recent literature, due to the controllable online prediction overhead, the model-free localization approach becomes more popular, especially after deep learning technique is widely adopted. For example, classical neural networks including restricted Boltzmann machine (RBM)  \cite{wang2017biloc,wang2015deepfi},  convolutional neural networks (CNN) \cite{chen2017confi},  deep residual networks (ResNet) \cite{wang2017resloc}, are utilized to extract CSI features and classified to different reference positions (RPs) with certain probability. By fusing the classification results together, the resultant localization accuracy
in terms of {\em median distance error} \cite{bevington1993data} can be improved, which ranges from 1.78m to 0.89m \cite{wang2017biloc,wang2015deepfi,chen2017confi,wang2017resloc}. Although the model-free based scheme requires relatively less computational resources, the localization accuracy is not satisfactory. In this paper, instead of using classification based mechanism as commonly adopted in the previous model or model-free based approaches, we propose to use a logistic regression \cite{peduzzi1996simulation} based scheme to directly model the continuous localization function, which is able to achieve sub-meter level accuracy (97.2cm medium distance error) in the standard laboratory environment under the single WiFi AP settings. We hope the proposed logistic regression based scheme can shed some light on the model-free localization technique and pave the way for the practical deployment of deep learning based WiFi localization systems. The main contributions of this paper are listed below.
\begin{itemize}
\item{\em Logistic Regression for Localization} In the localization problems, a straight forward idea is to compare the measured signal features in the operating stage with the collected features of RPs in the training stage and classify them according to pre-defined RPs. Although this approach restricts the candidate solution sets into several pre-determined locations and greatly reduces the computational resources for online feature extraction and comparison, the corresponding localization accuracy may be affected. Since the deep learning based approach has fast online prediction capability, we can trade off for the localization accuracy by directly modeling the signal feature versus location relations and utilizing continuous logistic regression.
\item{\em Unified Optimization Framework} In addition, we establish the general relationship between measured CSIs and the associated locations in order to propose a unified optimization framework for WiFi fingerprints based localization problems. Based on this framework, we shall be able to explain why the logistic regression based approach can achieve better localization accuracy than the traditional classification based mechanisms and how the proposed scheme can be further improved.

\item{\em Robust Training Processing} Moreover, to reduce the randomness induced by the temporal or spatial variations, we introduce a small perturbation in the training stage to improve the robustness. To be more specific, we collect CSIs from RPs and their neighboring areas and label them as CSIs at the corresponding RPs to improve the robustness of neural networks. We will show through numerical experiments that the proposed robust training process can improve about 30\%.
\end{itemize}

The rest of this paper is organized as follows. In Section~\ref{sect:pre}, we provide some preliminary information on the system model and CSI collection methods. The optimal localization formulation is discussed in Section~\ref{sect:prob} and the logistic regression based solution is proposed in Section~\ref{sect:sys}. We present our experimental results in Section~\ref{sect:experiment} and finally conclude this paper in Section~\ref{sect:conc}.

\section{Preliminaries} \label{sect:pre}
\begin{figure}
\centering
\includegraphics[width = 3.4 in]{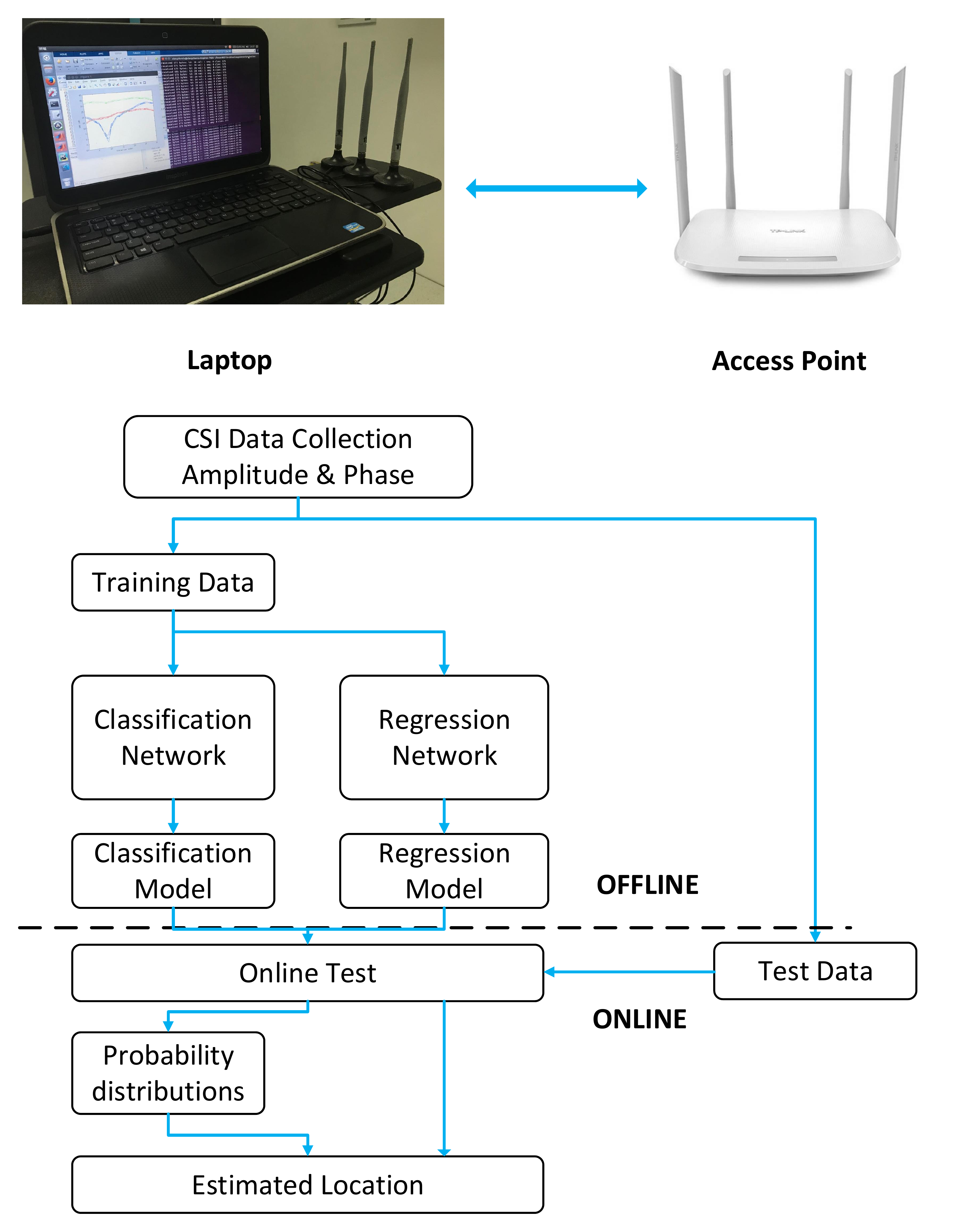}
\caption{The system architecture, the laptop collects CSI by receiving packets from the AP. The whole process can be divided into offline phase and online phase, the amplitude and phase of collected CSI data are used as training and test data.}
\label{fig:system}
\end{figure}

Consider a WiFi localization system as shown in Fig.~\ref{fig:system}, where the localization entity is a laptop equipped with $N_R$ receive antennas and Intel 5300 network interface card. Based on receiving the real time WiFi signals from an off-the-shelf AP with one transmit antenna, the localization entity applies the Linux 802.11n CSI Tool \cite{halperin2010predictable} and computes the CSI knowledge on a standard WiFi OFDM symbol basis with duration 3.2$\mu$s.

Without loss of generality, we denote $\mathbf{h}_{i}(\mathcal{L}, n) \in \mathbb{C}^{N_R}$ to be the collected channel state at the $i^{th}$ subcarrier and the $n^{th}$ OFDM symbol for the given location $\mathcal{L}$. By assembling CSIs from all $N_{sc}$ subcarriers, we have the overall CSI matrix for the $n^{th}$ OFDM symbol, $\mathbf{H}(\mathcal{L}, n) \in \mathbb{C}^{N_R \times N_{sc}}$, which is given by,
\begin{eqnarray}
\mathbf{H}(\mathcal{L}, n)  = \Big[
\mathbf{h}_{1}(\mathcal{L},n) \
\mathbf{h}_{2}(\mathcal{L},n) \
\cdots \
\mathbf{h}_{N_{sc}}(\mathcal{L},n) \Big],
\end{eqnarray}
and the localization function $g(\cdot)$ is therefore represented by,
\begin{eqnarray}
\mathcal{L} = g\left(\{\mathbf{H}(\mathcal{L}, n), \forall n \in [1,N] \}\right),
\end{eqnarray}
where $N$ denotes the number of OFDM symbols in each localization period. In the model based approach, we shall find the closed-form relationship between $\mathbf{H}(\mathcal{L}, n)$ and $\mathcal{L}$ by exploiting AOA and TOF features, which is in general very complicated. While in the model-free based scheme, we directly figure out the characteristics of the localization function $g(\cdot)$ and propose a better approximation by learning the collected CSIs and locations in the training stage.

The following assumptions are made through the rest of this paper. First, we ignore the imperfectness in the channel estimation process, e.g. the limited pilot power, the imperfect hardware components and additive noises. Second, to control the deployment complexity, we only collect the CSIs from some discrete RPs rather than sampling the entire fading environments. Last but not least, since the mathematical representation of {\em median distance error} is in general complicated, we use {\em mean distance error} (MDE) \cite{bevington1993data} instead as the performance measurement in the training stage as well as the loss function design.

\section{Problem Formulation}\label{sect:prob}

In this section, we formulate the localization problem using a generalized optimization framework. Denote $\hat{\mathcal{L}}_m$ and $\mathcal{L}_m$ to be the $m^{th}$ predicted and true locations respectively, and we can define MDE to be the average localization error over $M$ sampling positions, where the mathematical expression is given by $\frac{1}{M} \sum_{m=1}^{M} \|\hat{\mathcal{L}}_m - \mathcal{L}_m\|$. Therefore, we can describe the MDE minimization problem using the following optimization framework.
\begin{Prob}[MDE Minimization]
\label{prob:MDE_min}
\begin{eqnarray}
\underset{g (\cdot)}{\textrm{minimize}} && \frac{1}{M} \sum_{m=1}^{M} \|\hat{\mathcal{L}}_m - \mathcal{L}_m\| \\
\textrm{subject to} && \hat{\mathcal{L}}_{m} = g \left(\left\{\mathbf{H}(\mathcal{L}_{m},n_{m})\right\}\right), \\
&& \hat{\mathcal{L}}_{m}, \mathcal{L}_{m} \in \mathcal{A}, \forall m \label{eqn:const_1}
\end{eqnarray}
\end{Prob}
where $\mathcal{A}$ denotes the feasible localization areas and $\|\cdot\|$ denotes the vector norm operation as defined in \cite{peduzzi1996simulation}. In addition, we use notation $n_m \in [1, N]$ to represent the duration of the $m^{th}$ localization period with $N$ observed OFDM symbols.

Since the above minimization is over all the possible choices of functions $g(\cdot)$, traditional classification based approach aims to decompose the problem into two stages, where it computes the likelihood distribution with respect to several RPs in the first stage and simply applies a fusion technique to obtain the final result. Mathematically, this approach can be described as follows.
\begin{Prob}[Classification based Localization]
\label{prob:CBL}
\begin{eqnarray}
\underset{g_1 (\cdot), g_2 (\cdot)}{\textrm{minimize}} &&  \frac{1}{M} \sum_{m=1}^{M} \|\hat{\mathcal{L}}_m - \mathcal{L}_m\| \\
\textrm{subject to} && \hat{\mathcal{L}}_{m} = g_1 \left(\left\{\hat{\mathcal{L}}_{m} (n_m)  \right\}\right),\\
&& \hat{\mathcal{L}}_{m} (n_m) = \mathbf{p}_m^{T} (n_m)  \cdot \overline{\mathcal{L}_{RP}}, \\
&& \mathbf{p}_m (n_m) = g_2 \left(\mathbf{H}(\mathcal{L}_{m},n_{m}), \overline{\mathcal{L}_{RP}} \right), \\
&& \mathbf{p}_m (n_m) \in [0,1]^{N_{RP}},\\
&& \mathcal{L}_{m} \in \mathcal{A}, \forall m,
\end{eqnarray}
where $\overline{\mathcal{L}_{RP}}$ denotes the collected locations of all possible RPs, $\mathbf{p}_m (n_m)$ denotes the likelihood distribution with respect to $\overline{\mathcal{L}_{RP}}$ based on the $n_m^{th}$ OFDM symbol, and $N_{RP}$ denotes the number of RPs in the localization process\footnote{In the practical deployment, $M$ is often equal to $N_{RP}$ in order to reduce the testing and data processing complexity for modeling $g_2(\cdot)$ in the training stage.}.
\end{Prob}

In Problem~\ref{prob:CBL}, $g(\cdot)$ has been decomposed into functions $g_1(\cdot)$ and $g_2(\cdot)$, where the existing literatures focus on modeling $g_2(\cdot)$ as a typical classification problem and $g_1(\cdot)$ usually adopts the mathematical average operation or some Kalman filter \cite{cai2017cril} based techniques to fuse multiple observations together. Through this method, the searching space of potential locations has been greatly reduced, e.g. from all feasible locations defined by area $\mathcal{A}$ in Problem~\ref{prob:MDE_min} to $\overline{\mathcal{L}_{RP}}$ with dimension $N_{RP}$, which greatly reduces the computational complexity involved.

A straight forward question to ask is whether we can directly model the function $g^{\star}(\cdot)$ defined by,
\begin{eqnarray*}
g^{\star}(\cdot) = & \arg \min_{g(\cdot)} & \frac{1}{M} \sum_{m=1}^{M} \|\hat{\mathcal{L}}_m - \mathcal{L}_m\| \\
& \textrm{subject to} & \hat{\mathcal{L}}_{m} = g \left(\left\{\mathbf{H}(\mathcal{L}_{m},n_{m}) \right\}\right), \\
&& \hat{\mathcal{L}}_{m}, \mathcal{L}_{m} \in \mathcal{A}, \forall m,
\end{eqnarray*}
using the logistic regression concept\cite{peduzzi1996simulation}. With potentially larger optimization spaces, it shall be able to achieve better localization accuracy than the classification based scheme. To control the potential deployment complexity associated with the logistic regression scheme, we tighten the constraint \eqref{eqn:const_1} and assume that the original function $g^{\star}(\cdot)$ can be well approximated by $g^{\star}_{LR}(\cdot)$. Mathematically, we have
\begin{eqnarray*}
g^{\star}(\cdot) \approx g^{\star}_{LR}(\cdot) = & \arg \min_{g(\cdot)} & \frac{1}{M} \sum_{m=1}^{M} \|\hat{\mathcal{L}}_m - \mathcal{L}_m\| \\
& \textrm{subject to} & \hat{\mathcal{L}}_{m} = g \left(\left\{\mathbf{H}(\mathcal{L}_{m},n_{m}) \right\}\right), \\
&& \hat{\mathcal{L}}_{m}, \mathcal{L}_{m} \in \overline{\mathcal{L}_{RP}}, \forall m.
\end{eqnarray*}
It is worth to note that, theoretically, the above approximation can be improved when the number of RPs, $N_{RP}$, increases, which actually provides a meaningful tradeoff between the implementation complexity and the localization accuracy\footnote{When the number of RPs, $N_{RP}$, tends to infinity, we are able to characterize the function $g^{\star}(\cdot)$ in target area $\mathcal{A}$ with probability 1 by learning the function $g^{\star}_{LR}(\cdot)$. However, in the practical systems, we observe that the localization accuracy saturates when $N_{RP}$ exceeds some threshold value. }.

\section{Logistic Regression Solution} \label{sect:sys}
In this section, in order to find a better approximation of the function $g^{\star}_{LR} (\cdot)$, we consider to directly apply logistic regression based solution and the corresponding difficulties are obvious. First, with fluctuated wireless environments, the original sampled channel states contain random noises, which may significantly degrade the approximation accuracy offered by neural networks. Second, the design methodology for logistic regression based localization is unclear based on the existing literature. Last, the logistic regression based approach requires huge amount of data to train the neural network, which incurs significant overhead in the practical deployment. In this section, we introduce the proposed logistic regression based localization scheme in detail to address the above three challenges.

\subsection{Data Collection and Cleaning}
Instead of using generated data from numerical simulations, we collect the CSI conditions under 5.32GHz WIFi channel through a customized laptop with three receiving antennas as shown in Fig.~\ref{fig:system}. The training dataset contains 60000\footnote{To accelerate the model training, we install Keras on our server with Intel(R) Xeon(R) CPU E5-3680 and NVIDIA Tesla P100 GPU.} transmitted packets with length of 100 and the delay between packets is 4 ms. It means that various channel situations of 4 minutes in our experimental environment are logged in our dataset. For more convenient data processing, CSI data extracted by the Linux 802.11n CSI Tool is transformed into polar coordinates, i.e. $\mathbf{h}_{i}(\mathcal{L},n) = |\mathbf{h}_{i}(\mathcal{L},n)|e^{j\theta_{i}(\mathcal{L},n)}$, where $|\mathbf{h}_{i}(\mathcal{L},n)|$ and $\theta_{i}(\mathcal{L},n)$ denote the amplitude and phase information of $\mathbf{h}_{i}(\mathcal{L},n)$, respectively, and $j$ denotes the imaginary unit.

In the practical systems, the measured phase information, e.g. $\hat{\theta}_{i}(\mathcal{L},n)$ for subcarrier $i$, usually contains random jitters and noises due to the imperfect hardware components, which cannot be directly used for high accurate localization. To eliminate this effect, we apply
the classical phase calibration algorithm as proposed in \cite{xiao2012fifs}, and obtain,
\begin{eqnarray}
\theta_{i}(\mathcal{L},n) = \hat{\theta}_{i}(\mathcal{L},n) + \frac{ 2\pi i}{N_{FFT}}\delta - Z,
\end{eqnarray}
where $N_{FFT}$ denotes the size of Fast Fourier Transform (FFT) \footnote{Linux 802.11n CSI Tool is designed according to IEEE 802.11n protocol, and the FFT size is 64.}, $\delta$ denotes the time lag at the receiver, and $Z$ is the random measurement noise.

\subsection{Neural Network Architecture}
As the target of logistic regression is to find a better approximation of the non-convex function $g^{\star}_{LR}(\cdot)$, we choose the aggregated channel information, $\mathbf{H}(\mathcal{L}) = \big[\mathbf{H}(\mathcal{L}, 1),\ldots,$ $\mathbf{H}(\mathcal{L}, N)] \in \mathbb{C}^{N\times N_R \times N_{sc}}$, and the localization results, $\hat{\mathcal{L}}_{m} \in \mathbb{R}^{1\times2}$, to be the input and output matrices/vectors of neural networks, and select the loss function, $\mathbb{L}$, to be the original definition of $g^{\star}_{LR}(\cdot)$ given by\footnote{To simplify the system implementation, we choose the number of sampling positions, $M$, to be equal to the number of RPs, $N_{RP}$.},
\begin{eqnarray}
\mathbb{L} = \frac{1}{N_{RP}} \sum_{m=1}^{N_{RP}} \|\hat{\mathcal{L}}_m - \mathcal{L}_m\|.
\end{eqnarray}
Kindly note that this type of loss function is quite different from the traditional classification based approach, where the {\em cross entropy} \cite{de2005tutorial} is often applied to describe the difference between the classification results and the distribution of ground truth results. By minimizing the MDE with respect to the RPs, the logistic regression based approach can gradually converge to the non-convex function $g^{\star}_{LR}(\cdot)$ with satisfied performance via machine learning.

\begin{figure}
\centering
\includegraphics[width = 3.4 in]{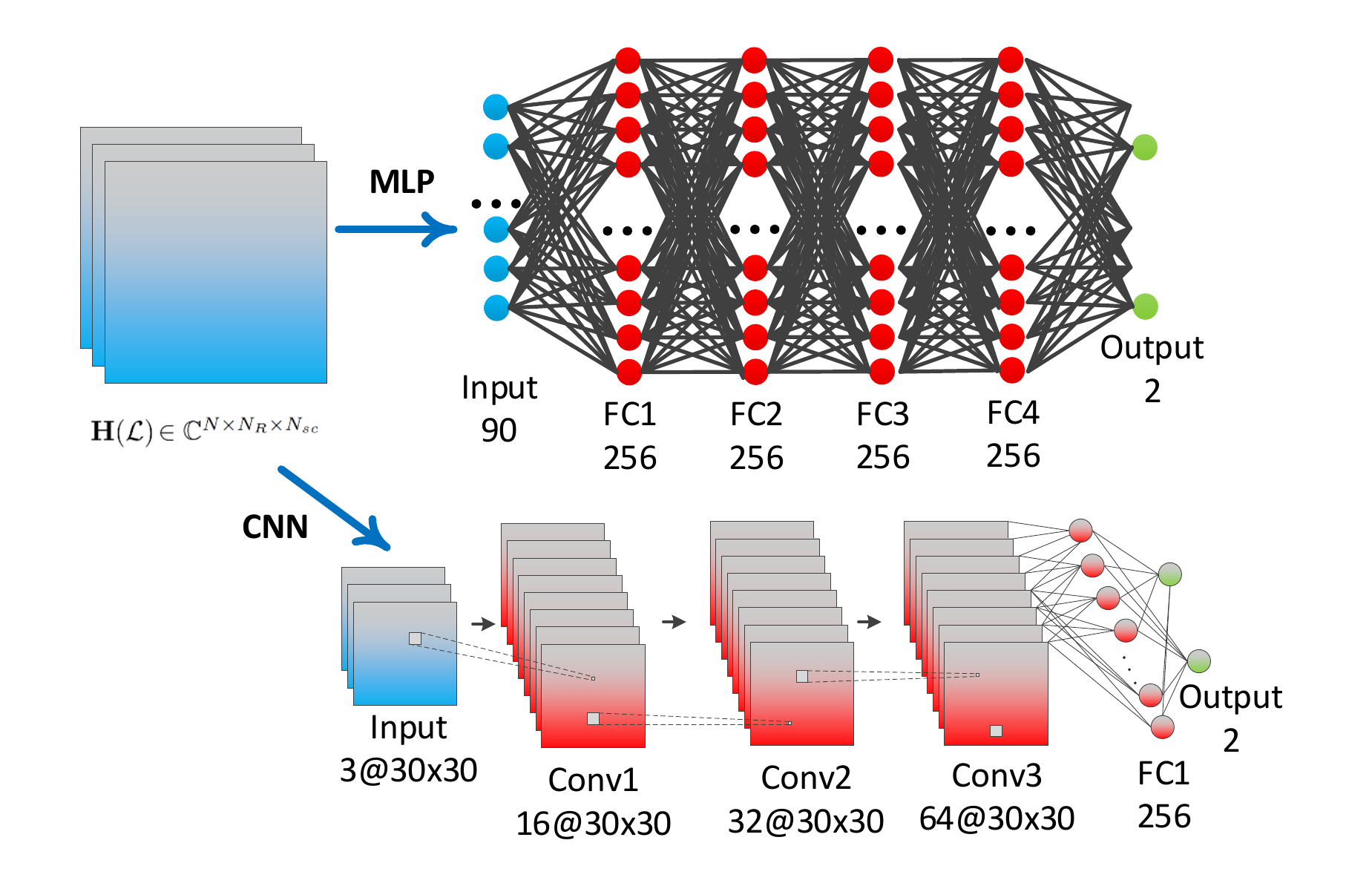}
\caption{The architecture of MLP and CNN. MLP is consisted of fully-connected (FC) layers and CNN also contains convolution layers, pooling layers besides that.}
\label{fig:network}
\end{figure}

In addition, to improve the representation and identification capability of neural networks, we exploit deeper structures of two classical neural networks, e.g., multi-layer perceptron (MLP) and convolutional neural network (CNN), as shown in Fig.~\ref{fig:network}. Compared with the fully connection structure of MLP networks, the convolution layers in CNN provide the feasibility to extract the features from CSIs across the spatial and frequency domains, which is more suitable for wireless fading environments as demonstrated in Section~\ref{sect:experiment}. To avoid the biasing effects of some unusual samples, we adopt the max pooling technique to get rid of unimportant features and apply dropout technique to further reduce the unimportant connections in the neural networks.  Meanwhile, to avoid the vanishing gradient problem, we also adopt rectified linear unit (ReLU) \cite{nair2010rectified} as the non-linear activation functions in each hidden layer. The detailed network configuration and parameters are summarized in Table~\ref{tab:parameter}, which is listed below.

\begin{table} [h]
\centering
\caption{An Overview Of Network Configuration and Parameters.}
\label{tab:parameter}
\footnotesize
\begin{tabular}{c c c}
\toprule
\textbf{Layers}&\textbf{MLP}&\textbf{CNN}\\
\midrule
Input Layer & $90$ & $3\times 30 \times 30$\\
\midrule
Hidden Layer 1 & FC 256 + ReLU & Conv $16\times3\times3$ + ReLU\\
\midrule
Hidden Layer 2 & FC 256 + ReLU& Conv $16\times3\times3$ + ReLU\\
\midrule
Hidden Layer 3  & FC 256 + ReLU & \tabincell{c}{Conv $16\times3\times3$ + ReLU  \\+ MaxPooling}\\
\midrule
Hidden Layer 4 & \tabincell{c}{FC 256 +  ReLU \\ + Dropout 0.3}& \tabincell{c}{FC 64 + ReLU\\ + Dropout 0.3}\\
\midrule
Output Layer & FC 2 + Linear & FC 2 + Linear\\
\midrule
Total No. of Para. & 225,282 & 235,682\\
\bottomrule
\end{tabular}
\end{table}

\subsection{Data Augmentation with Perturbation}
As mentioned in Section~\ref{sect:prob}, the logistic regression approach can be improved if $N_{RP}$ increases. However, in the practical deployment, the offline training stage requires careful measurement and labeling procedures, and the training time is in general proportional to $N_{RP}$. To control the deployment complexity, we introduce a perturbation based data augmentation scheme. A perturbation distance $\|\Delta \mathcal{L}\|$ is assumed to be much smaller than the localization distance $\|\hat{\mathcal{L}}_{m}\|$, i.e., $\|\Delta \mathcal{L}\| \ll \|\hat{\mathcal{L}}_{m}\|$. In this scheme, we have $\hat{\mathcal{L}}_{m} \approx \hat{\mathcal{L}}_{m} + \Delta \mathcal{L} = g \left(\left\{\mathbf{H}(\mathcal{L}_{m}+\Delta \mathcal{L},n_{m})\right\}\right)$, which means we can collect more CSI samples without changing the localization labels. Since the norm satisfies the triangle inequality, we have $\|\hat{\mathcal{L}}_m - \mathcal{L}_m\| -  \|\Delta\mathcal{L}\| \leq \|\hat{\mathcal{L}}_m - (\mathcal{L}_m + \Delta\mathcal{L})\| \leq \|\hat{\mathcal{L}}_m - \mathcal{L}_m\|+  \|\Delta\mathcal{L}\|$, and the MDE minimization can be rewritten as the following optimization problem.

\begin{Prob}[Logistic Regression with Augmentation]
\begin{eqnarray}
\underset{g (\cdot)}{\textrm{minimize}} &&\frac{1}{M} \sum_{m=1}^{M} \|\hat{\mathcal{L}}_m - (\mathcal{L}_m + \Delta\mathcal{L})\| \\
&& = \frac{1}{M} \sum_{m=1}^{M} \left(\|\hat{\mathcal{L}}_m - \mathcal{L}_m\|+\alpha \cdot \|\Delta\mathcal{L}\|\right) \\
\textrm{subject to}
&&\hat{\mathcal{L}}_{m}  \approx g \left(\left\{\mathbf{H}(\mathcal{L}_{m}+\Delta \mathcal{L},n_{m})\right\}\right), \\
&& \hat{\mathcal{L}}_{m}, \mathcal{L}_{m} \in \overline{\mathcal{L}_{RP}}, \forall m.
\end{eqnarray}
where $\alpha \in [-1, 1]$ represents a fine-tuning coefficient and can be determined in the training stage.
\end{Prob}

With the above formulation, we modify the loss function accordingly and use the augmented data sets in the training stage to improve the localization accuracy. The corresponding numerical results will be given in Section~\ref{sect:DA}.

\section{Experiment Results} \label{sect:experiment}

\begin{figure}[t]
\centering
\subfigure[Laboratory Scenario]{
\includegraphics[width=2.3 in]{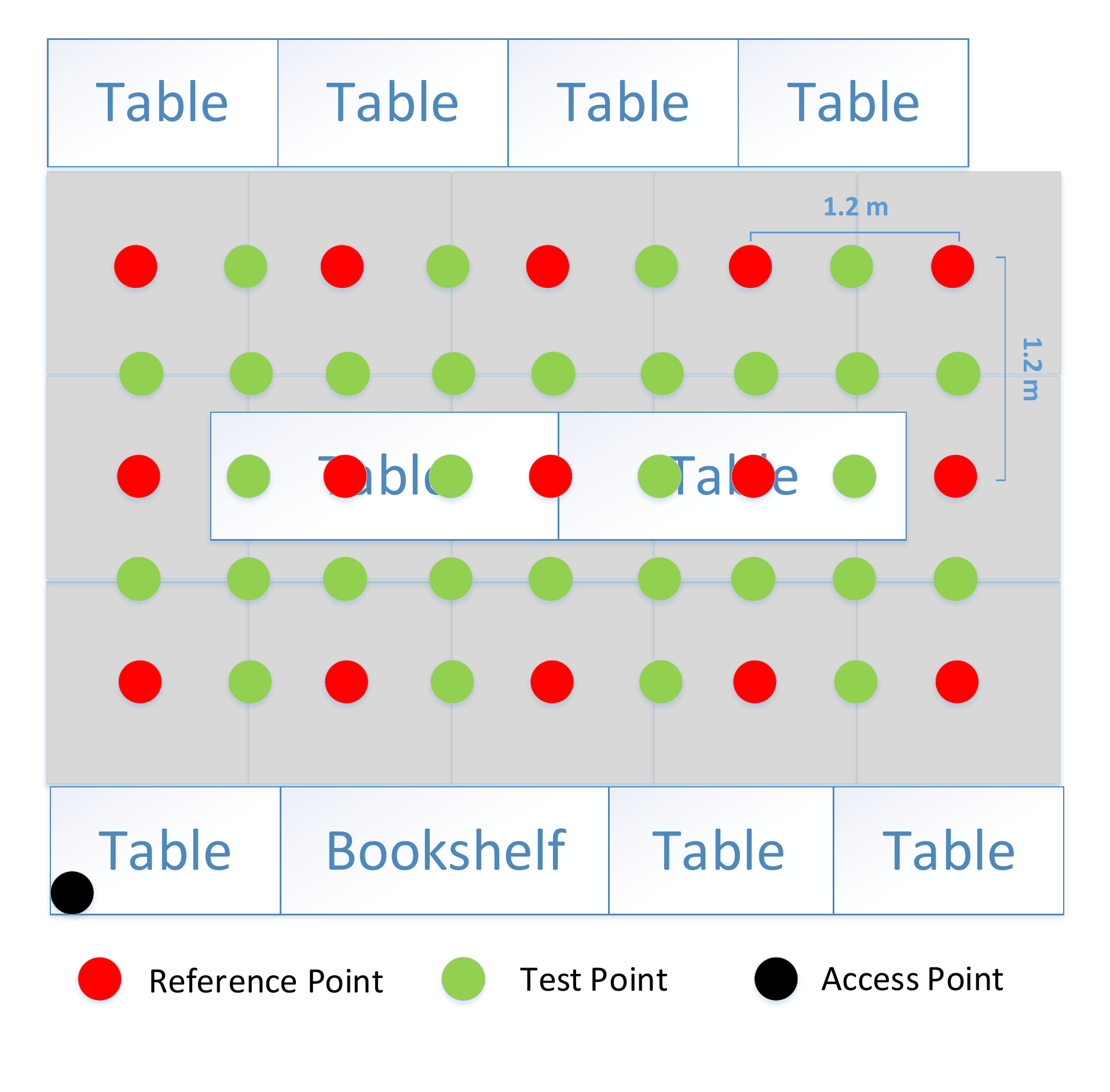}}
\hfill
\centering
\subfigure[Corridor Scenario]{
\includegraphics[width=3.4 in]{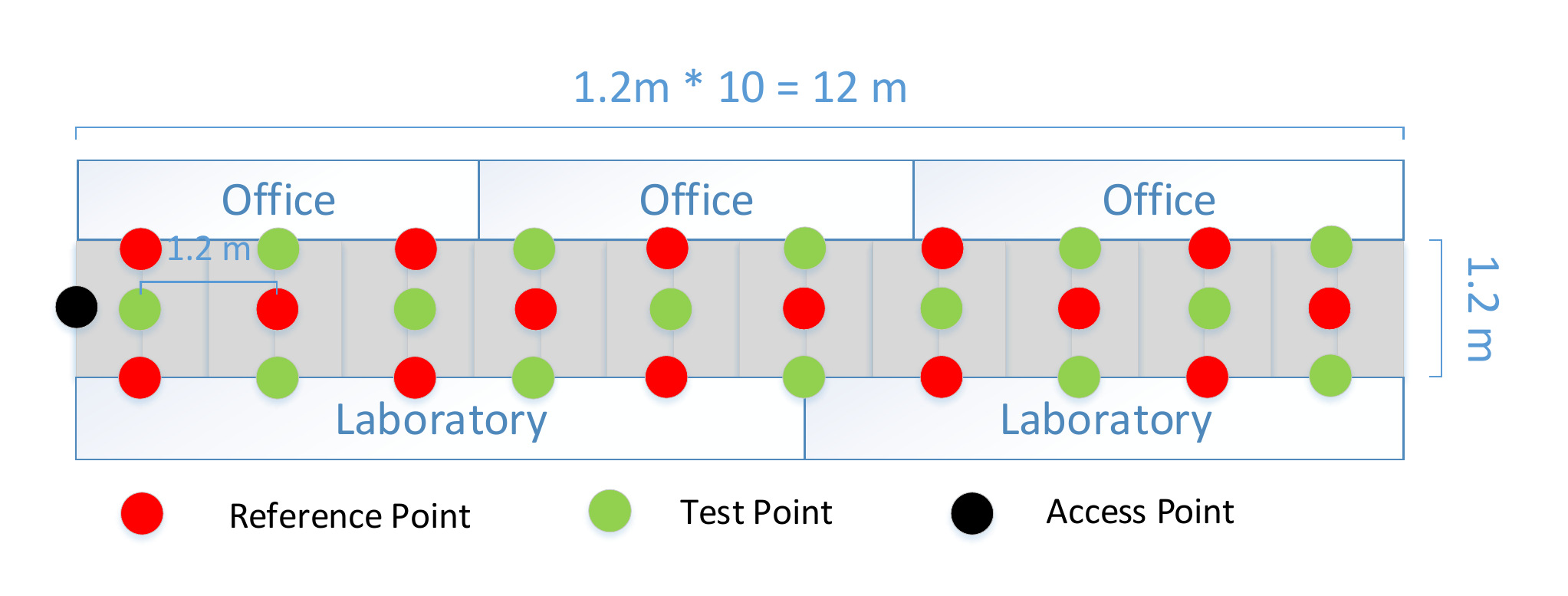}}
\caption{A sketch map of experiment environment, in which the red/ green/ black spots represent the location of the RPs, test points and AP, respectively. The distance between two adjacent RPs is 1.2m.}
\label{fig:environment}
\end{figure}

In this section, we provide some numerical results to show the effectiveness of the proposed logistic regression based approach for indoor localization. To be more specific, we compare the proposed scheme with two baseline systems, e.g., {\em Baseline 1}: KNN based localization and {\em Baseline 2}: classification based localization. We verify the proposed logistic regression based localization scheme in both laboratory and corridor environment, where the layout of testing scenarios are shown in Fig.~\ref{fig:environment}. With laboratory equipment, furniture, and people movements in the real situation, the tested wireless fading conditions cover most of the daily indoor scenarios with mixed LOS and NLOS paths.

\subsection{Logistic Regression vs. Classification}
\begin{figure}[t]
\centering
\subfigure[Laboratory Scenario]{
\includegraphics[width=3.4 in]{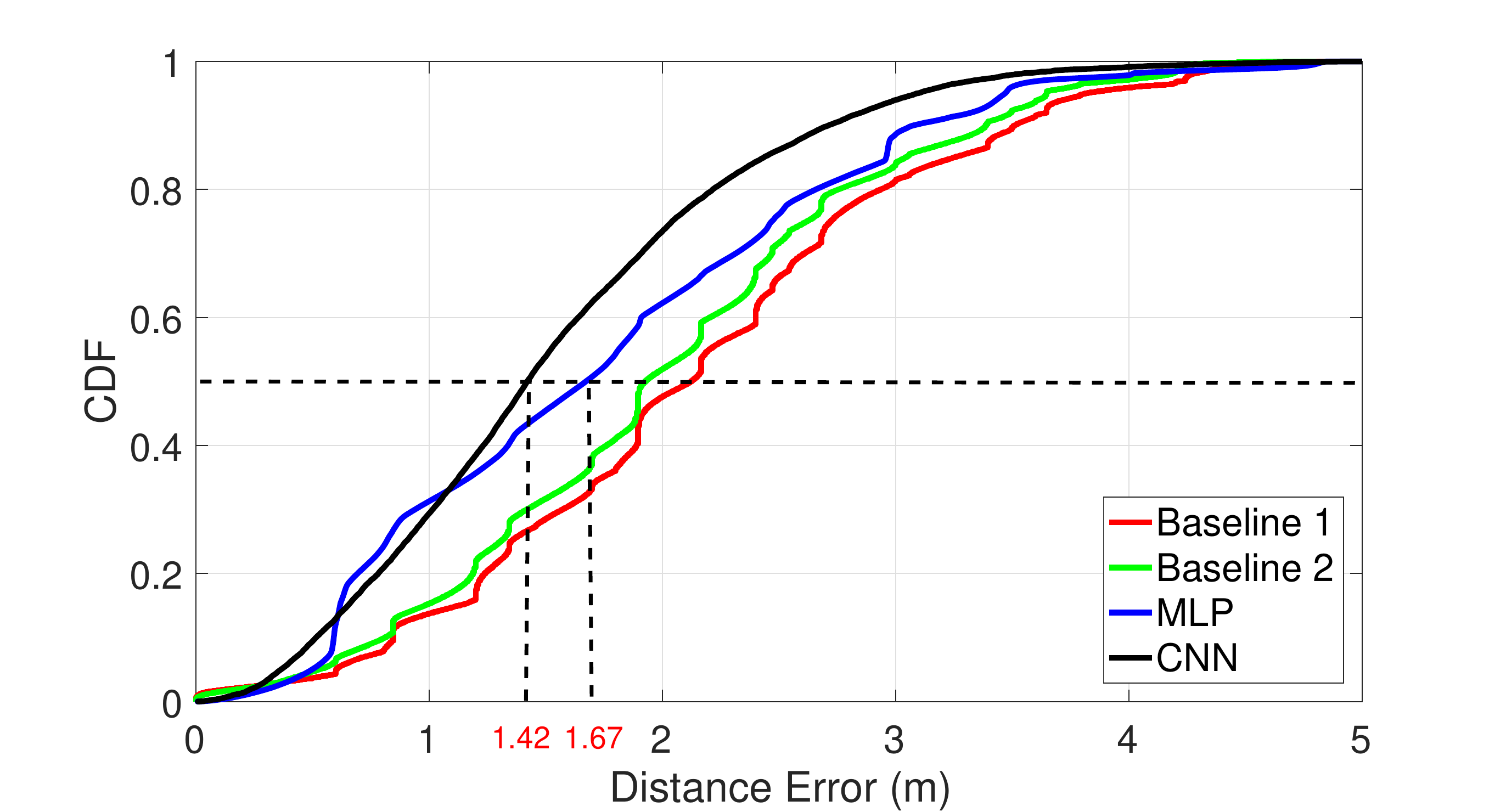}}
\hfill
\centering
\subfigure[Corridor Scenario]{
\includegraphics[width=3.4 in]{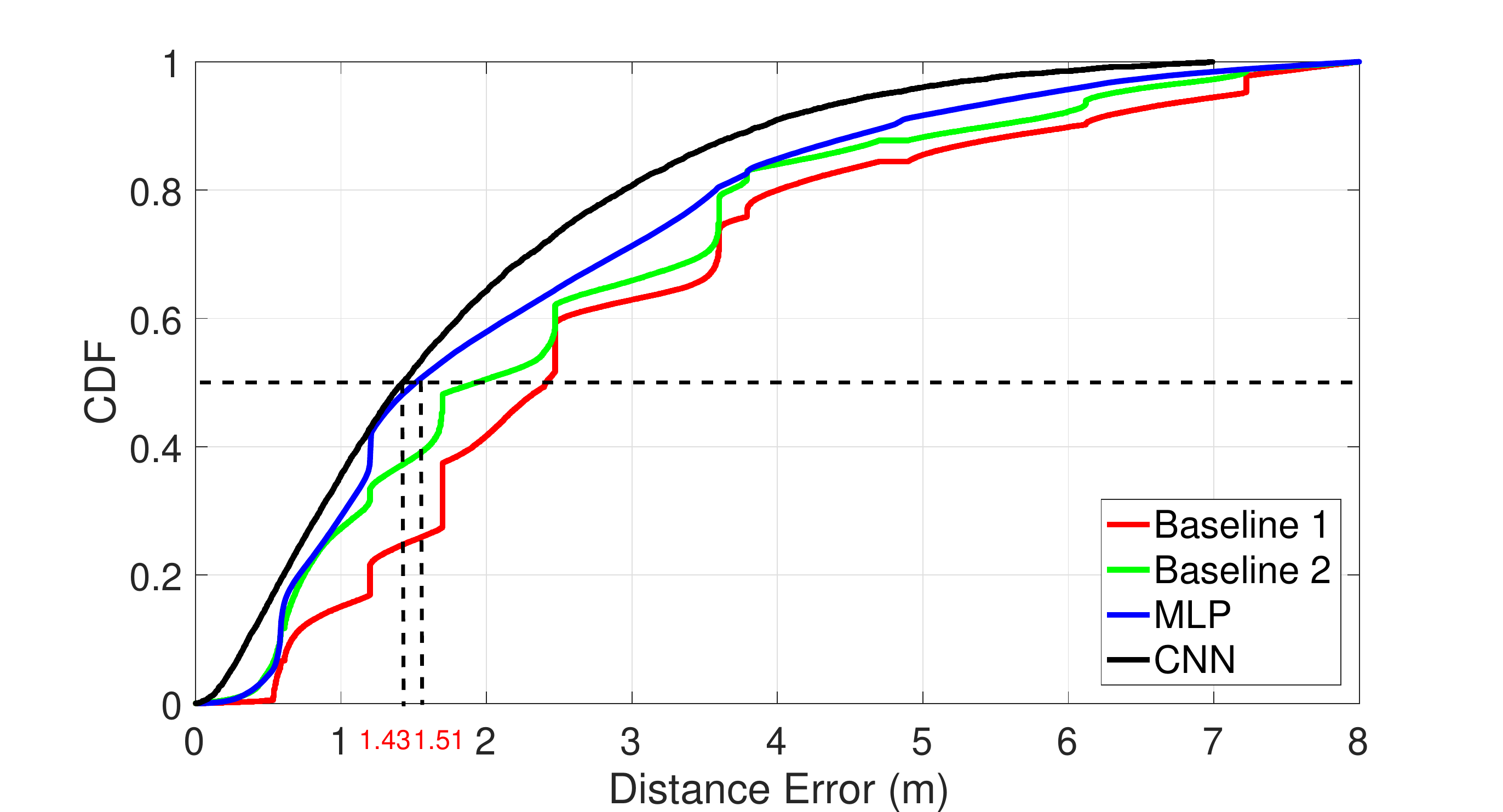}}
\caption{CDF of localization errors for different algorithms in the laboratory scenario and corridor scenario. The proposed regression approach with MLP and CNN architectures are compared with two baselines to test the location accuracy.}
\label{fig:Distance error}
\end{figure}

In the first experiment, we compared the proposed regression scheme with the above baselines by measuring the cumulative distribution function (CDF) of distance error in the laboratory as well as the corridor scenarios. Fig.~\ref{fig:Distance error} presents CDF of the localization distance error in the operating stage. The proposed regression based algorithms show superior localization accuracy over conventional algorithms, including KNN based localization ({\em Baseline 1}) and classification based localization ({\em Baseline 2}), for both two scenarios.

By comparing CNN-based approach (black solid curves) and MLP-based approach (blue solid curves), the former one achieves the median errors of 1.42m and 1.43m for the laboratory and corridor cases, respectively, which shows better positioning accuracy than the later one (1.67m for laboratory and 1.51m for corridor scenarios). This might due to the fact that CNN-based approach is able to capture the time domain correlations of multiple OFDM symbols, while MLP-based approach focuses on extracting the common features among all the observations.

\subsection{Effect of Data Augmentation} \label{sect:DA}
\begin{figure}
\centering
\includegraphics[width = 3.4 in]{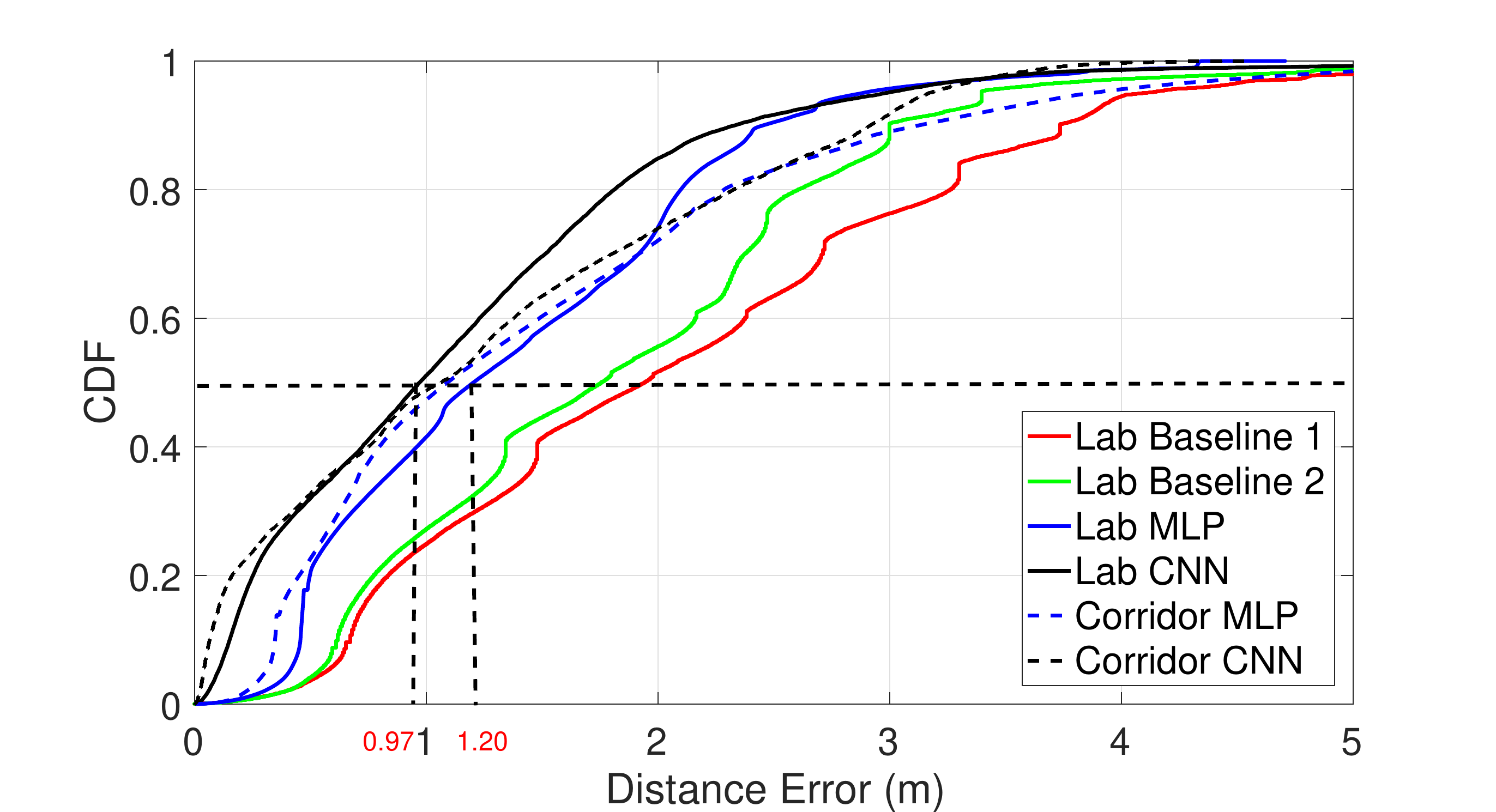}
\caption{CDF of localization errors for different algorithms. A new CSI dataset collected in laboratory environment are tested with MLP, CNN, {\em Baseline 1} and {\em Baseline 2} four different algorithms to verify the effectiveness of data augmentation.}
\label{fig:Augmentation}
\end{figure}

In order to verify the effectiveness of data augmentation with perturbation, we expand the original training data set by adding more perturbed samples with $\Delta \mathcal{L}$ less than 10cm. With the augmented data set, we re-train the neural networks and redo the same experiments in the operating stage. The CDF of distance error with different algorithms are compared in Fig.~\ref{fig:Augmentation}.

With the help of data augmentation, the localization accuracy of both CNN-based approach (black solid curves) and MLP-based approach (blue solid curves) has been improved, e.g., from 1.42m to 0.97m and from 1.67m to 1.20m for the median distance error in laboratory scenario, which corresponds to 32 and 28 percents improvement, respectively. In corridor scenario, the proposed data augmentation scheme shows similar localization accuracy improvement as well, which verifies the effectiveness as mentioned in Section~\ref{sect:sys}.

\section{Conclusion} \label{sect:conc}
In this paper, we propose a unified optimization framework for WiFi fingerprint based localization problems. Different from the conventional classification approach, we tackle the localization task via logistic regression based scheme and align the average localization error with the loss function design in neural networks to achieve better localization accuracy. Meanwhile, we propose data augmentation scheme using perturbations to expand the training data set and improve the localization accuracy. With all the above schemes, the localization system can achieve sub-meter level accuracy (97.2cm medium distance error) using a single WiFi AP.

\section*{Acknowledgement}
This work was supported by the National Natural Science Foundation of China (NSFC) Grants under No. 61701293 and No. 61871262, the National Science and Technology Major Project Grants under No. 2018ZX03001009, the Huawei Innovation Research Program (HIRP), and research funds from Shanghai Institute for Advanced Communication and Data Science (SICS).

\bibliographystyle{IEEEtran}
\bibliography{IEEEabrv,bb_rf}

\end{document}